\documentclass[aps,pre,twocolumn,notitlepage,nofootinbib,superscriptaddress]{revtex4}
\newif\ifpdf\ifx\pdfoutput\undefined\pdffalse\else\pdfoutput=1\pdftrue\fi

\usepackage{graphicx}% Include figure files
\newcommand{\avg}[1]{\langle{#1}\rangle}
\newcommand{\var}[1]{\mathrm{Var}\left({#1}\right)}

\newcommand{\beq}{\begin{equation}}
\newcommand{\eeq}{\end{equation}} 

\newcommand{\beqar}{\begin{eqnarray}}
\newcommand{\eeqar}{\end{eqnarray}} 
\newcommand{\ignore}[1]{}
\newcommand{\nignore}[1]{}
\newcommand{\WT}{\mathrm{WT}}
\newcommand{\OS}{\mathrm{SS}} 
\newcommand{\tmax}{t_{\rm{M}}}

\begin{document}
\title{Simulation and analysis of {\em in vitro} DNA evolution} 
\author{Morten Kloster}
\affiliation{Department of Physics, Princeton University, Princeton,
New Jersey 08544}
\affiliation{NEC Laboratories America, 4 Independence Way, Princeton,
NJ 08540}
\author{Chao Tang}
\email[Corresponding author: Chao Tang, NEC Laboratories America, 4
Independence Way, Princeton, NJ 08540. Tel: (609)-951-2644; Fax:
(609)-951-2496. E-mail: ]{tang@nec-labs.com}
\affiliation{NEC Laboratories America, 4 Independence Way, Princeton,
NJ 08540}
\affiliation{Center for Theoretical Biology, Peking University,
Beijing 100871, China}
 
\begin{abstract}
We study theoretically the {\it in vitro} evolution of a DNA sequence by
binding to a transcription factor. Using a simple model of protein-DNA 
binding and available binding constants for the {\it Mnt} protein, we 
perform large-scale, realistic simulations of evolution starting from a 
single DNA sequence. We identify different parameter regimes characterized
by distinct evolutionary behaviors. For each regime we find analytical 
estimates which agree well with simulation results. For small population
sizes, the DNA evolutional path is a random walk on a smooth landscape. 
While for large population sizes, the evolution dynamics can be well
described by a mean-field theory. We also study how the details of the
DNA-protein interaction affect the evolution.
\end{abstract}

\maketitle

The concept of evolution has not only fundamentally shaped our view of 
biology, but also found rich and profound applications in bioengineering 
and biotechnology. In particular, {\it in vitro} evolution has been widely
used to evolve DNA~\cite{Liu02}, RNA~\cite{RNA} and proteins~\cite{Arnold}.
In evolution, mutations at the 
molecular level are selected at the functional level. A quantitative 
theory of an evolutionary process would reqire a quantitative understanding
of the selection process (e.g. fitness function, landscape, selection 
pressure, etc.). While this is in general difficult to achieve for natural 
or laboratory evolution, there are simple cases where such a quantitative 
description is readily available. Based on experiments of RNA virus 
evolution~\cite{Holland}, Levine and 
colleagues~\cite{TsiLevKes,KesLevRidTsi,RidLevKes} studied a simple model
of evolutionary process in a smooth landscape in which the fitness of an
individual is given as the sum of many individual contributions that can
be mutated independently. Their studies found good agreement between theory
and simulations for small population sizes and for equilibrium mean field
theory, but the evolution dynamics turned out to be pathological in 
large population sizes where extremely rare mutations are exponentially 
amplified, yielding infinite speed of evolution. Peng 
{\em et al.}~\cite{PenGerHwaLev} proposed DNA binding to proteins as a
model system for evolution in a smooth landscape and studied a model where
a large population of long DNA molecules were subjected to high mutation 
rates and selected by how strongly they bind to a protein (the 
histone-octamer was mentioned as a possible example). Their model can be
well described by a continuum equation and they have shown that the average
distance to the highest affinity sequence exponentially approaches its 
equilibrium value.~\cite{PenGerHwaLev}
 
In a recent experiment, Dubertret {\it et 
al.}~\cite{Libchaber} studied the {\it in vitro} evolution of DNA sequences
via binding to the {\it lac} repressor protein which is kept unchanged. 
In the experiment, a round of evolution consists of amplification and 
mutation of the DNA pool by error-prone PCR (Polymerase Chain Reaction)
followed by selecting a fixed fraction of the DNA sequences via washing off
relatively weak protein binders. Starting from a mix of random DNA
sequences and by sequencing a number of DNA sequences after each round, 
they were able to observe the dynamics of DNA population evolving towards
the wild type (WT) sequence. In such a molecular breeding experiment, the
relation between the genotype (DNA sequence) and the phenotype (binding 
affinity to a protein) is direct and simple. If the binding constants are 
known for various DNA sequences, the selection process can be modeled 
quantitatively and it can then serve as a model system for quantitative 
analysis of molecular evolution. 

In this paper, we study theoretically the {\it in vitro} evolution of DNA
sequences via binding to the {\it mnt} repressor protein. The system of
DNA-{\it mnt} is perhaps the best experimentally characterized system of
sequence-specific DNA-protein binding,~\cite{Stormo97,Stormo98,Stormo01} 
and is particularly suited for a thorough quantitative study of molecular
evolution. Specifically, the binding constants of the WT DNA sequence and
many other sequences, including all the sequences one point-mutation away
from the WT, are measured experimentally.~\cite{Stormo97}
Furthermore, it has been demonstrated that the binding energy of a sequence
can be approximately decomposed as the sum of the contributions from the
individual bases.~\cite{Stormo97,Stormo01}
This {\it additive} form of binding energy greatly simplifies the 
analysis---it enables us to perform realistic large scale simulations as 
well as to obtain analytic solutions and estimates in various cases. 
In our study, we explore various regimes of experimentally accessible
parameters and we find very distinct evolution dynamics in different
regimes. 

\section{Models and Methods}
We assume that the binding energy between a DNA molecule and the {\it
mnt} protein is given by the sum of the contributions from individual
base pairs,
\begin{equation}
E(S)=\sum_i \epsilon_i(S_i), 
\label{energy}
\end{equation}
where $S_i\in A,C,G,T$ is the base at the $i$-th position of the DNA
sequence and $\epsilon_i(S_i)$ is the contribution to the binding energy
from the $i$-th position for which we use the experimentally determined
value in Ref.~\cite{Stormo97}. The relative binding constants are then
$K(S) = \prod_i K_i(S_i)=\prod_i e^{-\beta\epsilon_i(S_i)}$.~\footnote{In 
addition to this sequence-specific binding, DNA molecules may bind to the
transcription factor in a non-specific manner~\cite{NSB1,NSB2,NSB3}, 
$K_{tot}(S)=K(S)+K_{\rm{NSB}}$, and this binding dominates for DNA 
sequences that are very far from $\WT$. In this paper, we only consider
DNA sequences that are sufficiently close to $\WT$ so that the 
non-specific binding can be neglected---we still include it as a parameter
in our simulations, but ignore it in the analytical estimates.
See the appendix for more on non-specific binding.} We start
with a population size $N$ of DNA
molecules of the same sequence that is significantly different from the
WT, avoiding the potential problem of enrichment dominating the 
evolution~\cite{Libchaber} (i.e. sequences very close to the WT in the 
initial pool being amplified exponentially). An iteration of the
evolutionary process consists of an amplification with mutation followed
by a selection. During amplification the population is doubled $I$ times
(corresponding to, e.g. $I$ cycles of PCR), so that the population size
is increased to $2^IN$. We assume that at each duplication there is an
error rate of $r$ per base (see the appendix for more on this).
The population is then subject to a
selection process via binding to the {\it mnt} protein. Each DNA molecule
is selected with a probability $\frac{1}{1+e^{\beta(E(S)-\mu)}}$, where
the chemical potential $\mu$ is chosen such that the expected number of
selected DNA molecules is $N$. We iterate the evolutionary process
until at least $90\%$ of the DNA molecules are WT, and we denote the
number of iterations required $t_M$.

\noindent{\bf Simulation.} The binding site for the {\it mnt} repressor
consists of 17 important base pairs, at positions 3 through
19.~\footnote{The {\it mnt} repressor is a dimer of two identical
halves, which causes the $\WT$ to be palindromic around position 11. 
The actual wild type sequence differs from the sequence with the
highest affinity at one position (19), we denote the highest
affinity sequence by $\WT$ for convenience.~\cite{Stormo97}}
For our starting sequence, we chose, by random mutations from $\WT$, a
sequence that differs from $\WT$ at $m=6$ positions. We denoted by $\OS$
the starting sequence (Table~\ref{start}).
A simulation starts with $N$ copies of $\OS$. 
We call a specific sequence of mutations that take an $\OS$ to the $\WT$ 
an evolution path. Each path contains the six required mutations in some
order, and may contain additional mutations.~\footnote{If a molecule is
mutated at more than one position during a single duplication cycle, 
we randomly assign an order to these mutations. As a precaution, we have
also performed simulations where no more than one mutation per duplication
cycle per molecule was allowed, and verified that there were no qualitative 
differences.} There are $6!=720$ minimum
paths that only contain the six required mutations. We iterate the 
evolutionary process until at least $90\%$ of the DNA molecules are WT. 
We denote the number of iterations required $t_M$ and record the number
of molecules coming from each different path.

\begin{table}
\begin{tabular}{|c|c|c|c|c|c|c|}
\hline
        Position        & 4     & 7     & 9     & 10    & 13    & 15 \\
\hline
        $\WT$ base      & G     & C     & A     & C     & T     & G \\
\hline
        $\OS$ base      & T     & A     & C     & T     & C     & C \\
\hline
        $\Delta K$      & 2.13  & 10.0  & 5.0   & 5.5   & 7.2   & 8.3 \\
\hline
\end{tabular}
\caption{The starting sequence $\OS$. $\Delta K=e^{-\beta\Delta E}$ is the 
ratio of the binding constants due to the mutation at the specific position.}
\label{start}
\end{table}

\section{Results}
\begin{figure}
\includegraphics[width=3in]{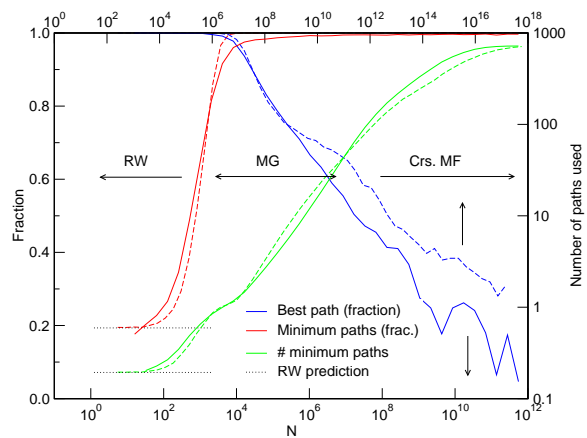}
\caption{The average fraction of $\WT$ contributed by the 
best path, the total fraction from all minimum paths, and the number of 
different minimum paths used, for different population sizes $N$. The
parameters $r=10^{-4}$ (solid lines), $r=10^{-7}$ (dashed lines), and
$I=1$ are held fixed. The various regimes (random walk, middle ground,
and crossover to mean field) are indicated.}
\label{regimes}
\end{figure}

Some of the key quantities from a single simulation run are the fraction
of $\WT$ that was produced through minimum paths, $f_{\rm{min}}^{\WT}$, 
the number of minimum paths used, $n_{\rm{min}}$, and the fraction of the
$\WT$ produced through the single best path in the simulation, 
$f_{\rm{best}}^{\WT}$. Fig.~\ref{regimes} shows how these quantities
depend on the population size (averaged over many simulations). We see 
that $f_{\rm{min}}^{\WT}$ is small for very small $N$ and is very close 
to 1 for large $N$, with a fairly sharp transition, whereas 
$f_{\rm{best}}^{\WT}$ slowly decreases from 1 with increasing $N$. This 
indicates that we may expect to find qualitatively different behavior for 
small and large population sizes.

\noindent{\bf Small $N$: Random Walk.}
Let us first consider the case of $N=1$ and $I=1$, i.e. a single DNA 
molecule is duplicated once and one of the two molecules is selected at 
each iteration. If there are no mutations during the duplication, the
two molecules are identical and nothing interesting can happen. If there
is a mutation of type\footnote{Mutation ``type" specifies both the position
and the bases involved.} $i$ (the chance of multiple mutations in the 
same duplication is negligible), the binding constant of the copy will 
be $\Delta K_i$ times that of the original, and the chance of selecting
the mutant is $\frac{\Delta K_i}{1+\Delta K_i}$, which is high for 
favorable mutations ($\Delta K_i>1$) and low for unfavorable ones.
The DNA molecule will thus perform a biased random walk: The molecule
will make a step whenever a mutant is selected, and the steps that improve
the binding to the protein are favored over the steps that reduce the
binding. Considering the probability of making each step, we find that we
have exactly a random walk in the energy landscape given by the binding
energy---in equilibrium, ${\rm Prob}(S)\propto e^{-\beta E(S)}$.

Now consider a population size $N>1$, keeping $I=1$. As long as $N$ is 
sufficiently small ($N \ll 1/r$), the chance of a mutation
happening in any single iteration is very low. When a mutation happens, 
it will almost certainly either ``die out'' (disappear from the 
population) or spread through the whole population before the next 
mutation occurs, so that most of the time the population consists of 
$N$ identical DNA molecules.
During selection, the chance of choosing a particular combination of DNA 
molecules is proportional to the product of their binding constants.
Thus, if there are $m$ mutants of type $i$ in the population at iteration $t$,
the chance that there will be $m+j$ mutants in the population the next 
iteration is $(\Delta K_i)^{2j}$ times the chance that there will be $m-j$
mutants (as we select exactly half the DNA, the combinatorics are identical).
The probabilities $p_{\pm}^N(m)$ that $m$ identical mutants in a 
population of size $N$ will spread through the whole population ($p_+$)
or will die out ($p_-$) then satisfy 
\beq
 p_+^N(N-m)=(\Delta K_i)^{2m} p_-^N(m).
\label{P_pm}
\eeq
%In particular, the chance that a mutant of type $i$ will spread through 
%the whole population ($p_+^N(1)$) is $(\Delta K_i)^{2(N-1)}$ times the 
%chance that the reverse process will happen ($p_-^N(N-1)$) (the reverse 
%process is that one DNA molecule of the original type---being then a 
%mutation---will spread through a population otherwise consisting entirely
%of the type $i$ mutants).
Including the probability that a mutant will be created and
selected in the first place, we find the rate $R_i$ at which a single
mutation of type $i$ will cause the whole population to be replaced:
%\beq
% R_i = \frac{Nr}{3}\cdot\frac{\Delta K_i}{\Delta K_i+1}p_+^N(1) = \frac{Nr}{3}\cdot\frac{(\Delta K_i)^{2N}-(\Delta K_i)^{2N-1}}{(\Delta K_i)^{2N}-1}
% \approx \left\{\begin{array}{cl}\frac{Nr}{3}\left(1-\frac{1}{\Delta K_i}\right) & , \Delta K_i>1 \\ 0 & , \Delta K_i<1 \end{array} \right.
%\eeq
\beqar
 R_i & = & \frac{Nr}{3}\cdot\frac{\Delta K_i}{\Delta K_i+1}p_+^N(1) = \frac{Nr}{3}\cdot\frac{(\Delta K_i)^{2N}-(\Delta K_i)^{2N-1}}{(\Delta K_i)^{2N}-1} \nonumber \\
 & \approx & \left\{\begin{array}{cl}\frac{Nr}{3}\left(1-\frac{1}{\Delta K_i}\right) & , \Delta K_i>1 \\ 0 & , \Delta K_i<1 \end{array} \right.
\eeqar
where Eq.~(\ref{P_pm}) and the condition $p_+^N(m)+p_-^N(m)=1$ are used,
and the approximation is valid for $|2N\log \Delta K_i|>>1$.
The population again describes a random walk,
but this time the energy landscape is $2N-1$ times the binding energy of
a single DNA molecule---in equilibrium, ${\rm Prob}(N\mbox{ copies of }S)
\propto e^{-\beta(2N-1)E(S)}$.

%As the binding energy change for a mutation is typically a few $k_B T$,
%even for very small population sizes this will be a completely biased
%random walk, i.e. a mutation that reduces the binding energy will not 
%survive. Correspondingly, a mutation that increases the binding energy is
%guaranteed to spread through the whole population once it makes up a 
%significant fraction of the population. This allows us to calculate the 
%probability that a given mutation will survive. As long as the mutants 
%make up a negligible fraction of the DNA population, the chemical 
%potentials for the selections will equal the binding energy for the
%original sequence. A mutation $i$ thus has a chance $\frac{\Delta 
%K_i}{1+\Delta K_i}$ of surviving a selection. If it survives, it will be
%duplicated, producing two identical ``children" that may or may not survive
%the next selection. As long as the fraction of mutants is negligible, the
%fates of the children are independent, and the probability that a mutation
%will die out is the probability that it will not be selected in the first
%place plus the probability that, though it was selected, both children
%eventually die out:
%\beq
%  p_0 = \frac{1}{1+\Delta K_i}+\frac{\Delta K_i}{1+\Delta K_i}p_0^2,
%\eeq
%which has the (nontrivial) solution $p_0=\frac{1}{\Delta K_i}$, i.e. the
%chance of survival is $1-\frac{1}{\Delta K_i}$.

\begin{figure}
\includegraphics[width=3in]{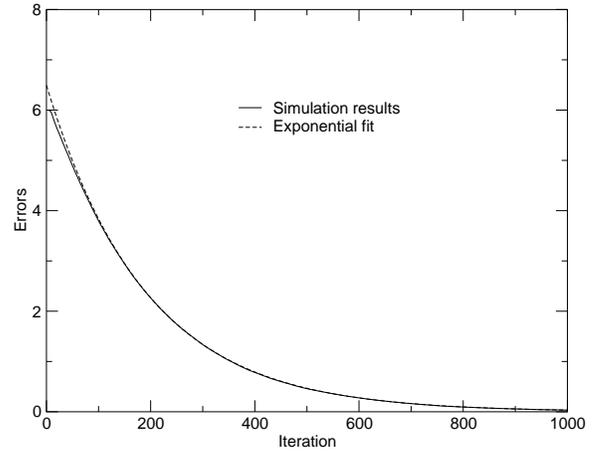}
\caption{Average number of different bases between DNA pool and WT as
function of time, and the exponential fit. $r=10^{-4}$, $N=256$, $I=1$, 
averaged over 8192 runs.}
\label{RWspeed}
\end{figure}

The average time needed to improve the DNA pool by one base relative to
$\WT$ can now be estimated as
\beq
  \avg{T} \approx \left[\sum_i \frac{Nr}{3}\left(1-\frac{1}{\Delta 
K_i}\right)\right]^{-1}+\frac{\log(N)}{\log(\frac{2\Delta K}{1+\Delta
K})},
\label{RWtimeeq}
\eeq
where $\Delta K$ is a typical value for $\Delta K_i$. The first term is
the time required to create a ``seed" mutation: The sum is over all possible
correct mutations, $\frac{Nr}{3}$ is the chance of each mutation occurring
in a given iteration, and $1-\frac{1}{\Delta K_i}$ is the chance that the
mutation survives. The second term is the time required for the mutation to
spread through the population (for each mutant, $\frac{2\Delta K_i}{1+\Delta
K_i}$ is the average number of surviving children after one iteration),
which is negligible for small $N$. Since the first term, which dominates, is
inversely proportional to the distance from WT (i.e. the number of terms in
the sum), the average speed of the DNA pool will be proportional to that
distance. Fig.~\ref{RWspeed} shows the distance from WT as a function of
time, and except in the beginning, it can be almost perfectly fitted to an
exponential---this is similar to the result in~\cite{PenGerHwaLev}, which is for
a very different regime. The corrections for the beginning are precicely what we would
expect from the second term: It reduces the speed when the speed is large
and causes a short delay. Our result for the evolution speed in the random
walk (RW) regime is very similar to the one found in~\cite{KesLevRidTsi} for
a birth-rate model: As long as the second term of equation~\ref{RWtimeeq} is
negligible, the evolution speed of the DNA population is proportional to the
mutation rate and to the population size $N$. 

\begin{figure}
\includegraphics[width=3in]{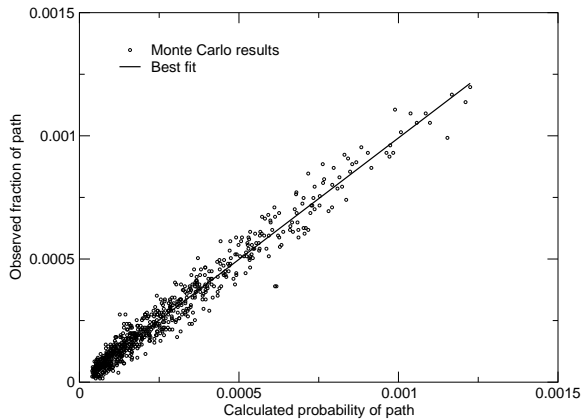}
\caption{Probabilities of individual minimum paths in random walk
regime---observed vs. predicted. $r=10^{-7}$, $N=512$, $I=1$, averaged over
131,072 runs.}
\label{RWprobs}
\end{figure}

Given a sequence $S$, the chance that mutation $i$ will be the next
surviving favorable mutation (in the limit of small $N r$ and large $N$)
is simply
\beq
  P_{RW\rm{mut}}(S,i) = \frac{1-\frac{1}{\Delta K_i}}{\sum_j
1-\frac{1}{\Delta 
K_j}},
\eeq
where the sum is over all possible favorable mutations to $S$. The
chance of a given path $\pi$ will thus be $P_{RW\rm{path}}(\pi) = \prod_n 
P_{RW\rm{mut}}(S_n^\pi,\pi_n)$. Fig.~\ref{RWprobs} compares these
predicted values with simulation results (which are Poisson distributed),
and as we can see, they agree very well---the total observed normalized 
variance is 745.2, vs. expected 720. The approach to the small $N r$
limit for the fraction of minimum paths out of all the paths used, is 
shown in Fig.~\ref{regimes}.

When $I>1$, the mutations with large $\Delta K$ will be significantly more
likely to happen first, as they are more likely to survive the now
stronger selection. The probability of each path can still be calculated 
(in the same limit as before), but this is far more complicated than for 
the case of $I=1$.

The random walk approach clearly doesn't work when the second term of 
equation~\ref{RWtimeeq} exceeds the first term: A mutation will not have
time to spread through the whole population before the next favorable 
mutation occurs.

\noindent{\bf Large $N$: Mean Field.}
For sufficiently large $N$, we expect mean field (MF) behaviour---there is
no significant difference between one experiment/simulation and another,
and each path contributes a fixed fraction to the total WT DNA produced. 
In principle, in the limit of $N\to\infty$ the fraction of population $S$
at time $t$, $f(S,t)$ can be traced from iteration to iteration and the
chemical potential determined from the selection criterion $\sum_S
f(S,t+1/2)/[1+exp(E(S)-\mu(t))]=2^{-I}$, where $f(S,t+1/2)$ is the
fraction of $S$ just after the amplification but before the selection.
This would require tracing all possible sequences---a tedious and often
impractical task. In practice, it suffices to restrict ourselves to a 
limited set of paths. With a fixed set of paths, we can find the
fraction of the DNA at each step of each path at each iteration, as well 
as the chemical potential used for each selection. The calculations proceed
the same way as for the finite $N$ simulations, except there is no 
randomness involved, and we discard the DNA that departs from the chosen 
paths.

From Fig.~\ref{regimes} we found that for large $N$, almost all the
$\WT$ is produced through minimum paths. Looking at the full parameter 
range we have explored, for $r=10^{-4}$ and $I=20$ the contribution to
$\WT$ through minimum paths is about 98\% for the largest $N$ we can 
simulate, and for all other parameters it's well above 99\% for large 
$N$---thus we expect any set that includes all the minimum paths to give a 
reasonable result. To get even more accurate results, we allow one
erroneous mutation (this already increases the number of paths from 720 
to 156,960) and verify that this is only a minor correction.

Let $n_0^{\OS,\pi}$ be the amount of $\WT$ produced from one molecule through
$\pi$, i.e.number of $\WT$ sequences at time $\tmax$
originating from a single molecule of sequence $\OS$ existing at the beginning 
of the experiment, through any specific path $\pi$.
Once we know all the chemical potentials $\mu(t)$ used for the selections,
we can use a set of recursion relations to find the average
$\avg{n_0^{\OS,\pi}}$ and the variance $\rm{Var}(n_0^{\OS,\pi})$,
as well as the probability
$P_+(n_0^{\OS,\pi})=P(n_0^{\OS,\pi}>0)$ that a single $\OS$ molecule at time
$t=0$ will yield some nonzero amount of $\WT$ at $t_M$ through $\pi$
(details in the appendix).

In the mean field regime, the amount of $\WT$ produced through each
minimum path should be relatively constant from one experiment to the next,
which gives us a lower bound for the population size:
\beq
  N>
\max_\pi{\frac{\rm{Var}(n_0^{\OS,\pi})}{\avg{n_0^{\OS,\pi}}^2}}.
\eeq
Table~\ref{boundaries} shows these bounds for a selection 
of parameters $I$ and $r$. The dependence on $r$ is roughly $N\sim r^{-m}$,
where $m=6$ is the number of different bases between $\OS$ and $\WT$.

\begin{table}
\begin{tabular}{|c||c|c|c|c|}
\hline
& \multicolumn{4}{c|}{$r$} \\ \hline
$I$ & $10^{-4}$ & $10^{-5}$ & $10^{-6}$ & $10^{-7}$ \\ \hline
1 & $1.1\cdot 10^{19}$ & $2.4\cdot 10^{24}$ & $7.5\cdot 10^{29}$ &
$2.9\cdot 
10^{35}$ \\
2 & $1.2\cdot 10^{19}$ & $2.9\cdot 10^{24}$ & $9.9\cdot 10^{29}$ &
$4.2\cdot 
10^{35}$ \\
4 & $1.6\cdot 10^{19}$ & $4.6\cdot 10^{24}$ & $1.8\cdot 10^{30}$ &
$9.0\cdot 
10^{35}$ \\
10 & $6.2\cdot 10^{19}$ & $3.8\cdot 10^{25}$ & $2.8\cdot 10^{31}$ &
$2.3\cdot 
10^{37}$ \\
15 & $1.4\cdot 10^{20}$ & $1.4\cdot 10^{26}$ & $1.4\cdot 10^{32}$ &
$1.4\cdot 
10^{38}$ \\
20 & $1.5\cdot 10^{19}$ & $1.5\cdot 10^{25}$ & $1.5\cdot 10^{31}$ &
$1.5\cdot 
10^{37}$ \\
\hline
\end{tabular}
\caption{Lower bounds for the mean field regime.}
\label{boundaries}
\end{table}

The speed of evolution in the mean field regime can be easily estimated in
the case of $I=1$. With $I=1$, half the population is removed during 
each selection, thus the chemical potential for the selection will be
close to the median binding energy in the population, and any DNA with
significantly higher binding energy than the majority will almost 
certainly survive. In the very first iteration of the evolution process, 
a fraction $r/3$ of the DNA will get each of the $m$ ``correct" mutations. 
In the following iterations, almost all of these improved DNA will survive
the selections, and the amount of improved DNA will thus roughly be doubled
in each iteration. After
\beq
  T_0\approx 1+\frac{\log(\frac{3}{mr})}{\log(2)}
  \label{MFspeedeq0}
\eeq
iterations, the improved DNA will have replaced the original population,
i.e. most of the DNA will only have $m-1$ errors relative to $\WT$.

Once the whole population has been improved by one base, the process is 
repeated. However, there have already been $T_0$ iteration in which the
improved DNA could improve further through mutations, and this even more
improved DNA has been amplified at the same rate as the regular improved 
DNA, i.e. the ``seed fraction" of improved DNA will now be 
$\frac{1}{2}\frac{T_0(m-1)r}{3}$ (the factor $\frac{1}{2}$ is because only
the copy gets mutated in each amplification). The time required to improve
the DNA pool by one base is thus roughly
\beq
  T(m^\prime) \approx 1+\frac{\log(\frac{3}{m^\prime T r})}{\log(2)},
  \label{MFspeedeq}
\eeq
where $m^\prime$ is the number of errors left. Corrections include
that some of the improved DNA will be lost during selection, reducing the
effective amplification from 2 to $\frac{2\Delta K}{\Delta K+1}$, and 
higher order corrections to the factor $T$ in the seed fraction. These and
other corrections can be addressed by considering an infinite length
model. \cite{Kloster}

\begin{figure}
\includegraphics[width=3in]{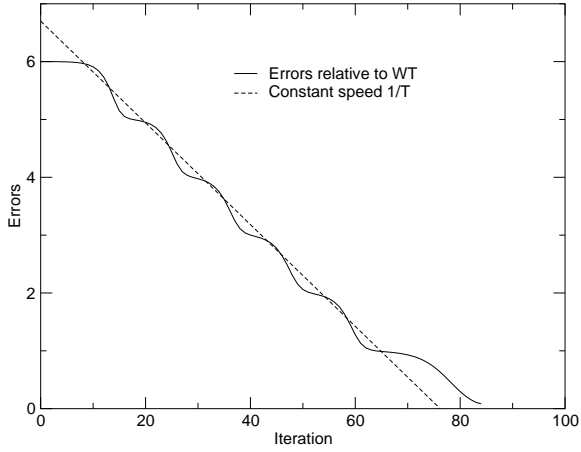}
\caption{Average number of different bases between DNA pool and $\WT$ as 
function of time. $r=10^{-4}$, $N=2^{39}\approx 5.5\cdot 10^{11}$,
$I=1$.}
\label{MFspeed}
\end{figure}

Fig.~\ref{MFspeed} shows how the average number of errors changes with
time, and we see that the evolution speed is almost constant---using $T(2)$
from equation~\ref{MFspeedeq} gives a very good fit. The first improvement
takes somewhat longer, as expected,
and so does the last improvement: For most of the DNA, the error at position
4 will be the last to be corrected, and the binding energy difference for 
that mutation is much smaller than for the others, thus the effective 
amplification is significantly smaller.

The effects of large $I$ and various analytical estimates for the mean field
regime are discussed in the appendix. 

\begin{figure}
\includegraphics[width=3in]{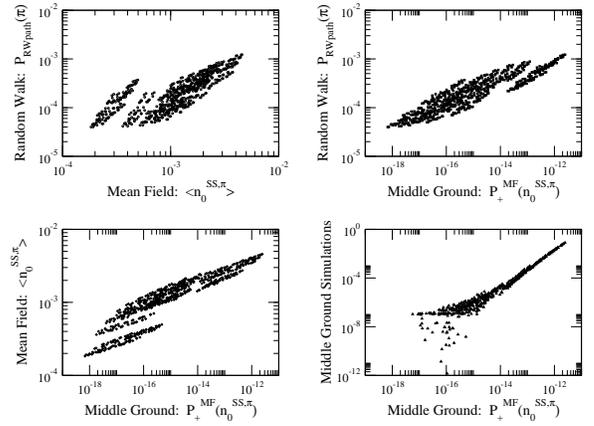}
\caption{The predicted average contribution from each path for the various
regimes, plotted against each other and, for the middle ground,
compared to simulation results (average of $2^{23}$ simulations with $I=1$,
$N=2^{24}$ and $r=10^{-7}$).}
\label{pathsvspaths}
\end{figure}

\begin{figure}
\includegraphics[width=3in]{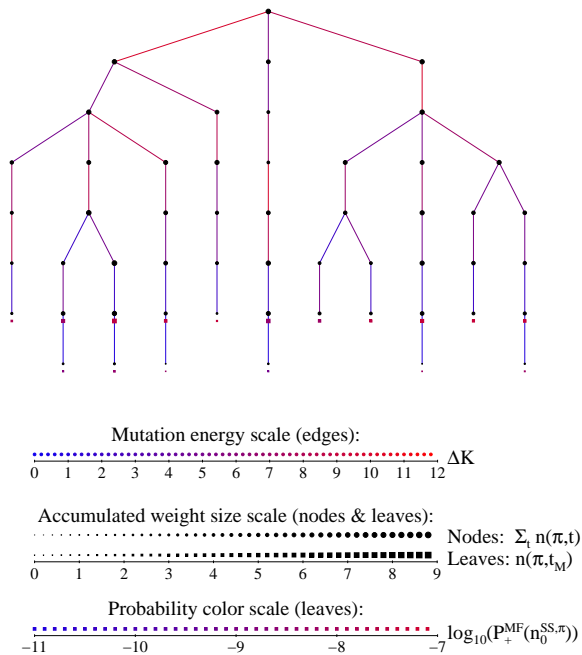}
\caption{Tree of evolution to WT (the root is $\OS$) for a sample simulation with
$N=2^{20}$, $r=10^{-4}$, and $I=1$. Node size shows the sum over all the 
iterations of the amount of the sequence (from that path) present during 
the simulation. Leaf size (square) shows the amount of $\WT$ produced (all
nodes annotated with a square are $\WT$). The color of the squares show 
the probability of that path using the mean field calculation.}
\label{tree}
\end{figure}

\noindent{\bf The Middle Ground.}
The argument used for the evolution speed in the mean field regime is 
qualitatively valid for all $N>\frac{1}{r}$ (the factor $T$ in the seed
fraction does not fully apply until $N>\frac{1}{r^2}$), thus we expect a
smooth transition from the RW evolution behavior, random with on average
exponential approach, to the mean field behavior, constant speed.
However, while all the key quantities shown in fig.~\ref{regimes}
vary smoothly with the population size, there is a significant
region where $f_{\rm{min}}^{\WT}\approx$1 and $f_{\rm{best}}^{\WT}>0.5$, i.e.
typically a single minimum path dominates. In this region the average contribution of
the different paths vary far more than in the RW and MF regimes
(Fig.~\ref{pathsvspaths}), and the region can be considered a third
parameter regime. Fig.~\ref{tree} shows the
result of a simulation in the middle ground regime, and while there in 
this simulation are 11 minimum paths that contribute $\WT$, 70\% of it 
comes from a single path. Note that all the minimum paths used here are
``probable" ones (fairly red).

As a single minimum path dominates in each simulation, the average contribution of
the different paths should depend strongly on how often that path dominates.
Given that the overall evolution behavior is similat to that of the mean field
regime, we can use as a first guess the probabilities
$P_+^{\rm{MF}}(n_0^{\OS,\pi})$ from the large $N$ discussion (we here add the
superscript $\rm{MF}$ to emphasize that these are mean field calculations).
Fig.~\ref{pathsvspaths} shows the results from simulations plotted against
this estimate, and there is a very accurate relationship between them
(it is not linear), at least for the most probable paths---for the less probable
paths, statistical errors are large. The middle ground region
corresponds fairly well to $\frac{1}{r}<N<(\sum_\pi 
P_+^{\rm{MF}}(n_0^{\OS,\pi}))^{-1}$, i.e. the regime ends approximately
when the population is large enough that, using the mean field chemical
potentials, we would expect to find at least some $\WT$ in most simulations.
There is a very large crossover region between the MF regime and the
middle ground, and a smaller region between the middle ground and the RW
regime (Fig.~\ref{regimes}).

Fig.~\ref{pathsvspaths} also shows the predictions
(or ``best guess" for the middle ground) for the various
regimes plotted against each other, and it is clear that they are very
different, though they are somewhat correlated.

\begin{figure}
\includegraphics[width=3in]{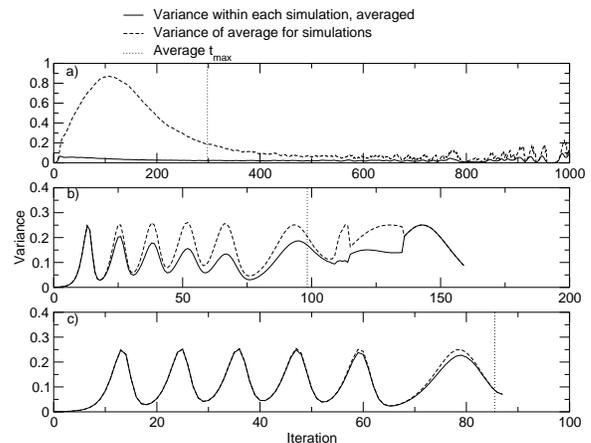}
\caption{The variance of the number of errors relative to $\WT$ as a
function of the time (iteration). $r=10^{-4}$ and $I=1$ for all graphs,
while $N=2^9$ for a), $N=2^{18}$ for b) and $N=2^{32}$ for c).}
\label{coherence}
\end{figure}

\noindent{\bf Experimental Signatures.}
It is difficult to completely and directly test the above theoretical 
analysis experimentally---one would need to sequence a large number of
DNA. A more practical choice is to consider only the distance from a 
sequence to $\WT$, i.e. the number of positions at which they differ, and
study its variance in different regimes. In the random walk regime, at 
most times the DNA pool consists of only a single sequence, thus the 
variance of this distance is, at any given time in any given run, almost
zero, but the variance from one run to the next can be very large. As we
increase the population size, the variance within a run increases somewhat,
but the variance between runs decreases drastically, and above the middle
ground we have almost perfect coherence (Fig.~\ref{coherence}).

\begin{figure}
\includegraphics[width=3in]{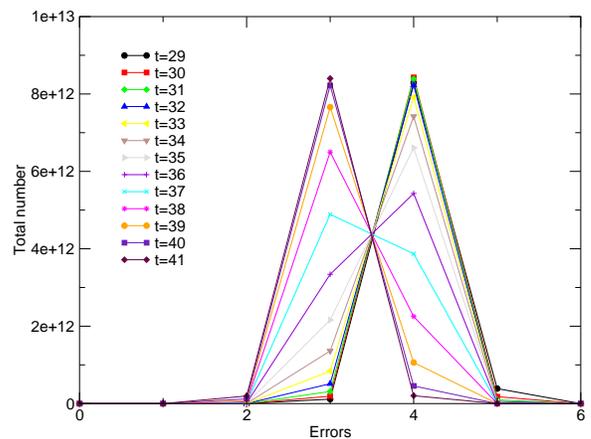}
\caption{Distribution of distances from $\WT$ as funtion of time.
$r=10^{-4}, I=1, N=2^{39}\approx 5.5\cdot 10^{11}$}
\label{distancedistr1}
\end{figure}

Fig.~\ref{distancedistr1} shows how the distribution of the distances
changes through (a part of) a simulation, and it confirms our earlier
assumptions: Only once the whole population has been ``upgraded" does 
further improvement start to become significant, i.e. there are at most 
two distances with significant population at any one time.

\section{Conclusion}
Our simulation and analysis show that in the simple case of additive
binding energy the evolution behavior of DNA-protein binding can be
understood quantitatively and rather completely. Depending on the
population size and the mutation rate, the evolutionary process exhibits
distinct behaviors in three parameter regimes. Our results are fairly
general as long as the potential is mainly additive and can be used to 
make sense of experimental data.

It is noteworthy that evolutionary processes via molecular breeding such 
as the one discussed here are fundamentally different from those where the
fitness of an individual depends solely on itself (e.g. on its genotype) 
and does not depend on others in the population. In the latter case 
the fitness landscape is
fixed~\cite{TsiLevKes,KesLevRidTsi,RidLevKes,Peliti,WooHig} and the more 
fit the better, while in the former case the fitness landcape is 
dynamic~\cite{PenGerHwaLev} 
and you just have to be better than the average---to be even better does 
not increase your fitness. The two cases can lead to, for example, very 
different evolution dynamics for large population sizes~\cite{TsiLevKes}.

The additivity of the binding energy gives rise to a smooth landscape,
which greatly simplifies the analysis. The inclusion of a small perturbative 
non-additive part to the potential would not change the picture, but would
nonetheless provide insights to the cases of more general potentials and
fitness functions.

We thank Qi Ouyang and Terry Hwa for very helpful discussions.

\section{Appendix}

\subsection{Approximations}
\noindent{\bf Mutation rates.}
In the main paper we assumed that all mutation rates were the same regardless
of which base (A,C,G or T) was mutated and which base it mutated into, while
in reality these rates can be very
different~\cite{MutRateRef1,MutRateRef2,MutRateRef3}. While this
assumption was done to simplify the analysis, the mutations rates will depend
on the specific experimental conditions under which the PCRs are performed,
and it might thus not have sufficed to choose just one set of base dependent
mutation rates.

The simulations, the mean field recursion relations and the analysis
of the random walk regime can all easily be altered to include
different mutation rates. Note that the exact results for the equilibrium
distributions in the random walk regime only hold (in the correct limit)
when mutation rates are the same for
opposite mutations. The evolution speed for the mean field regime is also
still essentially correct, although the least likely mutations will typically
occur last and will take longer, as the seed fraction is smaller (this is
also addressed in the infinite length model~\cite{Kloster}).

\noindent{\bf Non-specific binding.}
In the main paper we always started out with a DNA sequence that binds strongly to the 
protein, such that favorable mutations can be selected quickly. If we start out 
with a very poor sequence, or, in the simulations, increase the non-specific 
binding strength of the protein, then even highly favorable mutations will be 
only weakly selected, as non-specific binding will dominate.

\begin{figure}
\includegraphics[width=3in]{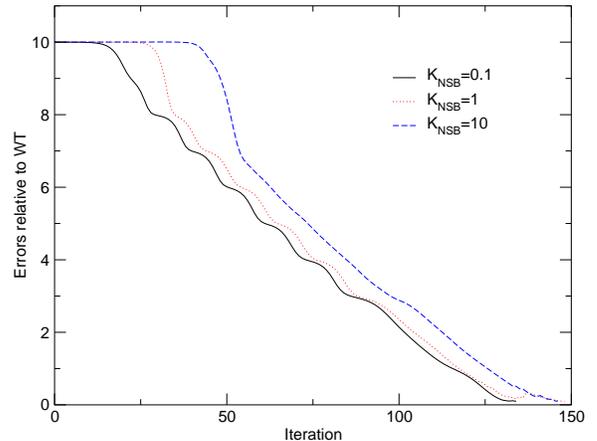}
\caption{Average distance from $\WT$ as a funtion of time for various values of 
non-specific binding. $r=10^{-4}, I=1, N=2^{33}\approx 8.6\cdot 10^{9}$}
\label{manymut}
\end{figure}

Figure~\ref{manymut} shows how the average number of errors changes through the 
iterations when we start with a sequence with 10 errors (the usual ones and 4 
more) and various values for the non-specific binding strength. For high 
non-specific binding strengths, none of the mutations are selected strongly, and
``nothing happens" until, by chance, several very good mutations combine, 
yielding a sequence that binds significantly more strongly than by non-specific 
binding alone - the DNA pool then abruptly improves by all those mutations at 
once, and the process then proceeds as usual. Also, the original sequence now 
contains several (3) very weak errors, and once only these remain, the ``one 
mutation at a time" assumption fails.

Note that a very large population size is required in order to be certain that a 
sufficiently good sequence will be produced (exponentially increasing the 
further into the non-specific binding regime we are), and the random walk regime 
no longer exists - for small $N$, most experiments would never generate the 
$\WT$ sequence.

\subsection{More on Mean Field}
\noindent{\bf Mean Field Recursion Relations.}
Here we derive a set of recursion equations in the MF regime. We assume 
that the number of iterations $\tmax$ and the chemical potentials 
$\mu(t)$ are known from a mean field simulation, as discussed in the
main paper.

Let $n_t^{S,\pi}$ be the number of $\WT$ sequences at time $\tmax$
originating from a single molecule of sequence $S$ existing a time $t$ 
in the experiment, through a specific path $\pi$ (which takes $S$ to WT).
Let $P_+(n_t^{S,\pi})$ be the chance that the molecule produces some 
nonzero amount of $\WT$ through $\pi$, $\avg{n_t^{S,\pi}}$ be the 
expected amount, and $\rm{Var}(n_t^{S,\pi})=\avg{(n_t^{S,\pi})^2}-
\avg{n_t^{S,\pi}}^2$ be the variance.

Given these quantities for time $t'$ right after a selection, we can
compute the 
values for time $t$ right before the selection:

\beqar
P_+(n_{t}^{S,\pi}) & = & p_{t,S} P_+(n_{t'}^{S,\pi})
\label{PosProbSelEq}\\
\avg{n_{t}^{S,\pi}} & = & p_{t,S} \avg{n_{t'}^{S,\pi}}
\label{AvgSelEq}\\
\avg{(n_{t}^{S,\pi})^2} & = & p_{t,S} \avg{(n_{t'}^{S,\pi})^2} 
\label{VarSelEq}\\
p_{t,S} & = & \frac{1}{1+e^{E(S)-\mu(t)}} \label{SelProbEq}
\eeqar

Similarly, given the quantities for $t'$ right after a cycle of PCR, we
can compute the values for time $t$ right before the PCR:
\beqar
P_0(n_{t}^{S,\pi}) & = & P_0(n_{t'}^{S,\pi})\left[1-(1-r)^L 
P_+(n_{t'}^{S,\pi})\right. \nonumber \\
& & \left. -\frac{r}{3}(1-r)^{L-1} P_+(n_{t'}^{S',\pi'})\right] 
\label{MFPCRProbEq}\\
\avg{n_{t}^{S,\pi}} & = & \avg{n_{t'}^{S,\pi}} + (1-r)^L
\avg{n_{t'}^{S,\pi}} 
\nonumber \\
& & +\frac{r}{3}(1-r)^{L-1} \avg{n_{t'}^{S',\pi'}},
\label{MFPCRAvgEq}
\eeqar
where $S'$ is the sequence reached by performing the first mutation in
path $\pi$ on $S$, $\pi'$ is the remaining path, and $P_0(x)=1-P_+(x)$.

The first term of equation~\ref{MFPCRProbEq}, $P_0(n_{t'}^{S,\pi})$, is
the chance that the original, unchanged DNA molecule does not produce any
$\WT$ at time $\tmax$ (through path $\pi$). The second term is the 
corresponding chance for the possibly mutated copy: If it was not mutated
(chance $(1-r)^L$), the chance of producing $\WT$ through path $\pi$ is 
$P_+(n_{t'}^{S,\pi})$. If it was mutated exactly the right way, i.e. 
according to the first step of path $\pi$ (chance $\frac{r}{3}(1-r)^{L-1}$),
the chance of producing $\WT$ through the remaining part $\pi^\prime$ of 
path $\pi$ is $P_+(n_{t'}^{S,\pi^\prime})$, while any other mutation would 
mean it is off the right path.

Equation~\ref{MFPCRAvgEq} is straightforward: $\avg{n_{t'}^{S,\pi}}$ is
the contribution from the unchanged molecule, and the other terms are for the
possibly mutated copy. The equation for $\avg{(n_t^{S,\pi})^2}$ follows
immediately. These equations cover the case where we only allow single 
mutations; there are additional terms when we allow multiple mutations 
(i.e. multiple steps along the path).

\begin{figure}
\includegraphics[width=6in]{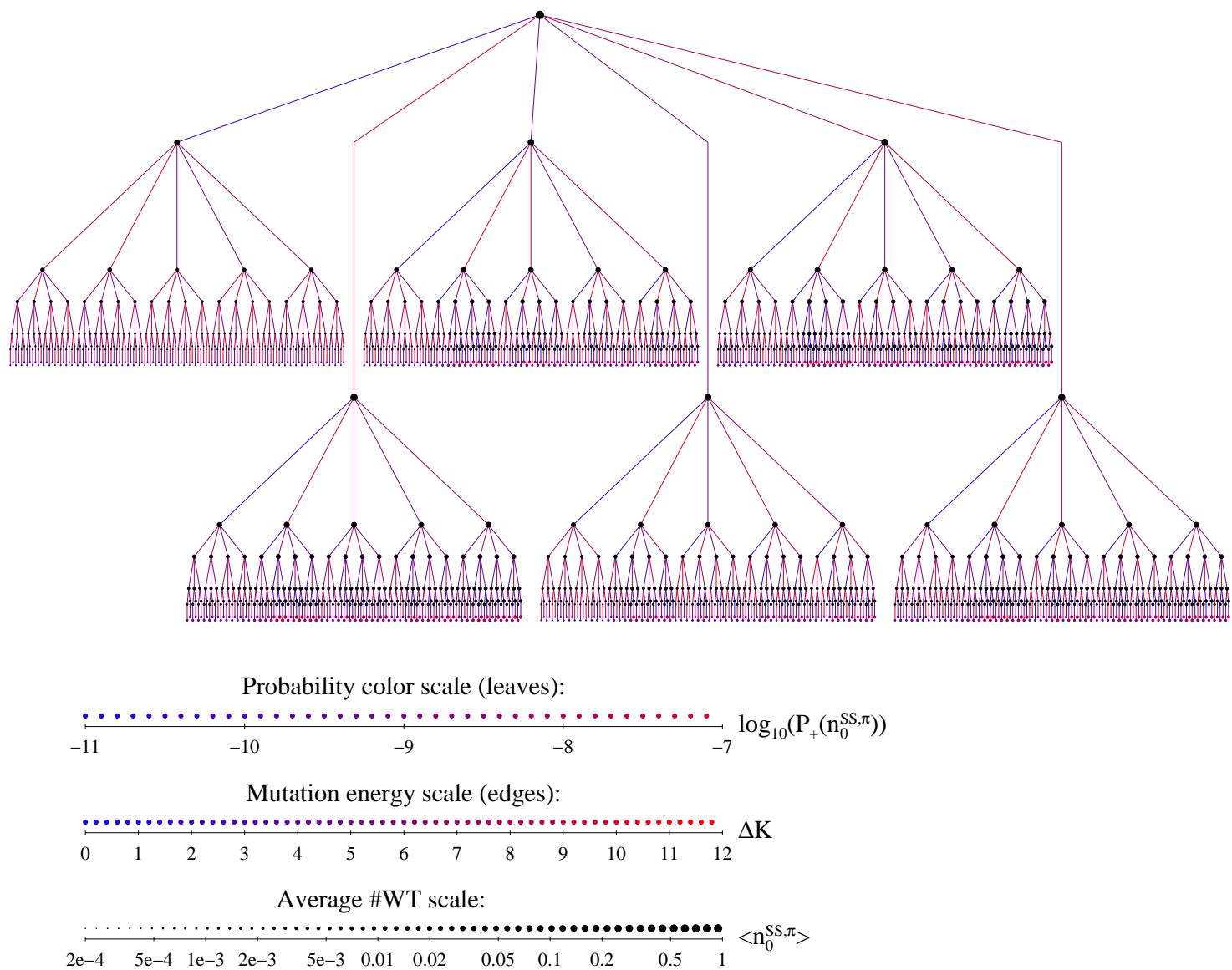}
\caption{Tree of evolution to $\WT$ for mean field regime. The root is
$\OS$, the leaves are $\WT$ produced through individual paths, the edges are
mutations. $r=10^{-4}$, $I=1$. Only minimum paths are included. Size of 
internal nodes is given by the sum of the size of the leaves.}
\label{MFtree}
\end{figure}

We trivially know that 
$P_+(n_{\tmax}^{\WT,-})=\avg{n_{\tmax}^{\WT,-}}=\avg{(n_{\tmax}^{\WT,-})^2}=1$, 
and the values are zero for all other sequences at that time. By
applying the above equations for all selections and PCRs in the experiment,
in reverse order, we find the desired quantities for time $t=0$. The results
of this calculation for $r=10^{-4}$ and $I=1$ are shown in Fig.~\ref{MFtree}.

\noindent{\bf Large $I$.}
As discussed for $I=1$, evolution works by first producing a seed fraction of 
improved DNA which is then amplified through subsequent iterations. Unlike the 
$I=1$ case, we can no longer consider all improved DNA (with a given number of 
errors) to be equivalent:

The maximum amplification (in one iteration) for DNA that is better than the 
majority of the population is $2^I$. However, for this amplification to actually 
occur, the binding constant of the DNA that is to be amplified must be at least 
$2^I$ times that of the typical DNA in the population. For $2^I\gg \Delta K$, 
the amplification will simply be $\Delta K$ per iteration.

The highest possible amplification in our model is thus 
$\frac{K(WT)}{K(OS)}\approx 2^{15}$. For very small $I$ we expect the number of 
iterations required to improve by one base (equations 7 and 8), and thus the total number of iterations $\tmax$, to be 
approximately inversely proportional to $I$ (this 
holds for all $I$ for the infinite length model~\cite{Kloster}). As we increase $I$, the various 
amplifications are gradually saturated (approach $\Delta K$), and for $I>15$ 
they are all saturated. Increasing $I$ beyond this only serves to increase the 
effective rate of mutation. Table~\ref{IterationNumbersTable} shows how 
$\tmax$ depends on $I$, and this matches our expectations well.

\begin{table}[htb]
\begin{tabular}{|c||c|c|c|c|c|c|c|c|c|c|}
\hline
$I$ & 1 & 2 & 4 & 6 & 10 & 15 & 20 & 30 & 50 & 100\\ \hline
$\tmax$ & 151 & 80 & 46 & 36 & 28 & 25 & 22 & 22 & 21 & 20\\
\hline
\end{tabular}
\caption{Number of iterations for mean field simulations. $r=10^{-7}$ for all 
values.}
\label{IterationNumbersTable}
\end{table}

For large $I$, the amplification of a given DNA sequence in an iteration is 
approximately the ratio of its binding constant to that of the typical DNA in 
the population, thus DNA that is much better than the average will be amplified 
far more strongly. In particular, DNA that mutates to $\WT$ during the very 
first iteration will receive the maximum possible amplification, and this 
process is the one that contributes the most $\WT$ in a simulation.

We can estimate this contribution and the corrections due to DNA that does not 
quite reach $\WT$ the first iteration, as well as the variation from this 
process. These calculations are carried out below and give us an 
estimate for the lower bound of the mean field regime for large $I$:
\beq
  N>\max_\pi \frac{\var{n_{\WT}^\pi (\tmax)}}{\avg{n_{\WT}^\pi (\tmax)}^2} 
\propto \left(\frac{3}{Ir}\right)^m.
\eeq

\begin{figure}
\includegraphics[width=3in]{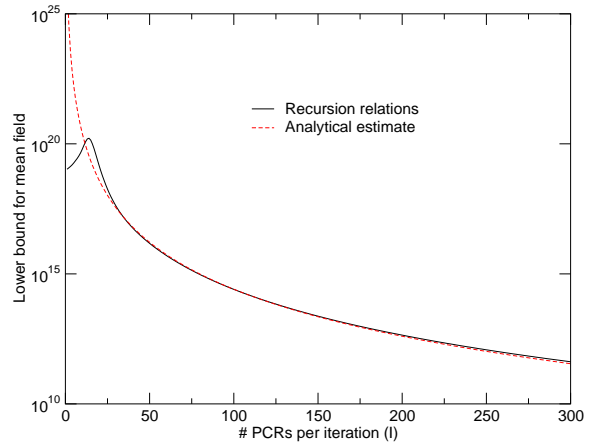}
\caption{Lower bound for mean field regime calculated from recursion relations 
vs. analytical estimate, as function of $I$. $r=10^{-4}.$}
\label{MFbounds}
\end{figure}

Figure~\ref{MFbounds} compares this estimate (equation~\ref{MFboundest}) to the 
(highly accurate) values found from the recursion relations 
(eqs.~\ref{PosProbSelEq}-\ref{MFPCRAvgEq}).

\noindent{\bf Mean field estimates: Large $I$.}
When $I$ is sufficiently large, the chemical potential for the selections is 
much lower than even the binding energy of $\WT$, thus the chance that a DNA 
molecule will survive a selection is proportional to the binding constant of the 
sequence. The chance that each initial error will be corrected during any given 
iteration ($I$ cycles of PCR) is $\frac{Ir}{3}$, but for each iteration that 
mutation $i$ has not happened, the DNA will be amplified by a factor $\Delta 
K_i$ less than those that have reached $\WT$. The number of $\WT$ molecules 
created (through all paths), as a function of time, is approximately
\beq
  \avg{n_{\WT}(t)}\approx f_I(t) 2^{It-m}\left(\frac{Ir}{3}\right)^m
    \prod_i \left(\frac{1}{1-(\Delta K^i)^{-1}}\right), \label{MFLargeIAvg}
\eeq
where $i$ runs over the mutated positions, and $f_I(t)$ is a function describing 
the ratio of $\WT$ molecules that would have survived all the selections so far. 
The powers of 2 is the total number of PCRs minus the number of cycles in which 
the DNA mutated (ignoring simultaneous mutations), and the last factor corrects 
for mutations that didn't happen the first iteration.

The main source of variance is whether or not a starting molecule produces a 
$\WT$ molecule that survives the first selection:
\beqar
  \lefteqn{\var{n_{\WT}^\pi(\tmax)} \approx \avg{(n_{\WT}^\pi(\tmax))^2}} 
\nonumber \\
  & \approx & \left(2^{I(\tmax-1)}\frac{f_I(\tmax)}{f_I(1)}\right)^2
    \frac{2^{I-m}}{m!}\left(\frac{Ir}{3}\right)^m f_I(1)  \label{MFLargeIVar} 
\\
  &\approx & \frac{2^m}{m!} \left(\frac{3}{Ir}\right)^m \prod_i 
\frac{\left(1-(\Delta K^i)^{-1}\right)^2}{\Delta K^i} \label{MFLargeIVar2}
\eeqar
where we have used $f_I(1)=\frac{\prod_i \Delta K^i}{2^I}$ ---this assumes that, 
before the selection for the first iteration, most of the DNA sequences have not 
been mutated, which fixes the chemical potential $\mu_1$. The first factor of 
eq.~\ref{MFLargeIVar} is the average number of $\WT$ molecules that a $\WT$ 
molecule that survives the first selection would produce by time $\tmax$, 
squared, and the remaining factors are the chance that such a molecule is 
produced; the expected number of $\WT$ molecules produced through a single path 
during the $I$ cycles of PCR in the first iteration times the chance $f_I(1)$ 
that each $\WT$ molecule survives the first selection.

We eliminated $f_I(\tmax)$ from equation~\ref{MFLargeIVar} by using 
equation~\ref{MFLargeIAvg}, setting $\avg{n_{\WT}(\tmax)}\approx 1$. From this 
we can get a rough estimate for the lower bound of the mean field regime by 
using $\avg{n_{\WT}^\pi(\tmax)}\approx \frac{1}{m!}$, as for $I=1$. However, for 
large $I$ we can do better by calculating the relative contributions of 
individual paths directly:
\beqar
  \avg{n_{\WT}^\pi(t)} & \approx & f_I(t) 2^{It-m}\left(\frac{Ir}{3}\right)^m
    g_m(\pi) \Delta K(\pi) \label{NWTPath} \\
  g_i(\pi) & = & \left(\frac{1}{i!}+\sum_{j=1}^{i-1} 
\frac{g_j(\pi_j)}{(i-j)!}\right)
    \frac{1}{\Delta K(\pi)-1},
\eeqar
where $\pi_j$ means the last $j$ steps of path $\pi$, and $\Delta K(\pi)$ is the 
total change in binding constant, i.e. the product of the $\Delta K$ for the 
individual mutations of path $\pi$. Knowledge of initial sequence is implicit.

Using equation~\ref{NWTPath} to eliminate $f_I(t)$ in 
equation~\ref{MFLargeIVar}, we find the bound for the MF regime directly:
\beq
  N>\max_\pi \frac{2^m}{m!} \left(\frac{3}{Ir}\right)^m \frac{\Delta 
K(\pi)}{(g_m(\pi))^2}
  \label{MFboundest}
\eeq

\noindent{\bf Mean field estimates: $I=1$.}
A DNA molecule that has higher binding energy than the average will typically 
survive a selection, and vice versa. The simplest estimate we can use for the 
behavior of the average binding energy is a linear change from the binding 
energy of $\OS$ to that of $\WT)$. 

Given that a DNA molecule mutates from $\OS$ to $\WT$ during the $\tmax$ 
iterations of the simulation, we estimate that it will survive all the 
selections iff the number of good mutations is always at least 
$\frac{mt}{\tmax}$. The number of ways this can happen is approximately 
$m\frac{(\tmax)^{m-1}}{m}$: There must be a mutation at the first iteration, 
giving the factor $m$ and leaving approximately $(\tmax)^{m-1}$ combinations for 
the remaining mutations. Given $m$ points randomly distributed on a circle of 
length m, there is with probability 1 exactly one point for which, when 
traveling clockwise from that point, you will always have visited more points 
than the length you have traveled (assuming $\tmax\gg m$, this continuum case 
should be good enough) --- thus, the chance that we started from the right 
point/mutation is $\frac{1}{m}$.

From this, the number of $\WT$ molecules produced from a single $\OS$ molecule 
in $\tmax$ iterations is about
\beq
  \avg{n_{\WT}(\tmax)}\approx 2^{\tmax-m}\left(\frac{r}{3}\right)^m 
(\tmax)^{m-1}, \label{MFAvgApprox}
\eeq
where the powers of two are the number of iterations where the DNA didn't change 
(and thus was duplicated).

The simulation ends when there are about as many $\WT$ molecules as there were 
$\OS$ molecules at the beginning:
\beqar
  1 & \approx & \avg{n_{\WT}(\tmax)}  \nonumber \\
  & \Downarrow & \label{tmax}\\
  \tmax & \approx & m+m\log_2{\left(\frac{3}{MR}\right)}-(m-1)\log_2(\tmax). 
\nonumber
\eeqar
This is very similar to $T_0+\sum_{m^\prime=1}^{m-1}T(m^\prime)$, using 
equations 7 and 8 and $\log_2(\tmax)\approx 
m\log_2(T)$ --- most of the discrepancy corresponds to the higher order 
corrections $\frac{-1}{\log(2)}$ found for the infinite length model~\cite{Kloster}.

Regarding the variance $\var{n_{\WT}^\pi(\tmax)}$ of the number of $\WT$ 
molecules produced from a single $\OS$ molecule through a single path $\pi$, any 
duplication that occurs before the $\WT$ is reached will only yield a factor of 
2 ---- the two molecules must independently develop to the $\WT$ --- while 
duplication once the $\WT$ is reached gives a factor of 4. Because of this, the 
variance is dominated by the extremely rare events where all the good mutations 
happen very early (the first $m+\Delta$ iterations), and the $\WT$ molecules are 
then simply duplicated until the simulation ends:
\beqar
  \lefteqn{\var{n_{\WT}^\pi(\tmax)}\approx \avg{n_{\WT}^\pi(\tmax)}^2} \nonumber 
\\
  & \approx & \sum_{\Delta\geq 0} 2^{2\tmax-2m-\Delta}
    \left(\frac{r}{3}\right)^m \frac{\left(m-2+\Delta\right)!}{\Delta!(m-2)!} 
\label{MFVarApprox1}\\
  & = & 2^{2\tmax-m-1}\left(\frac{r}{3}\right)^m. \label{MFVarApprox2}
\eeqar

If the DNA is mutated to $\WT$ in the first $m+\Delta$ iterations ($\Delta$ is 
the number of ``idle" iterations), it will then be duplicated $\tmax-m-\Delta$ 
times, giving a contribution $2^{2\tmax-2m-2\Delta}$. Also, the DNA will be 
duplicated during each idle iteration, but this merely increases the chance that 
a molecule will reach $\WT$, and thus gives a factor $2^\Delta$.

The number of ways the mutations can happen is 
$\frac{\left(m-2+\Delta\right)!}{\Delta!(m-2)!}$, assuming that there must be a 
mutation the first iteration --- we have ignored the possibility of multiple 
mutations in one cycle of PCR, which is a significant correction. 
Equation~\ref{MFVarApprox2} follows by using
\beq
  \sum_{\Delta\geq 0} x^{-\Delta}\frac{\left(n+\Delta\right)!}{\Delta!n!}
  = \left(\sum_{n\geq 0} x^{-n}\right)^{n+1}.
\eeq
We can then use the average $\avg{n_{\WT}^\pi(\tmax)}_\pi=\frac{1}{m!}$ to 
estimate (very roughly) the lower bound for the mean field regime.

For $r=10^{-4}$ we estimate $\tmax\approx65$ and thus 
$\var{n_{\WT}^\pi(\tmax)}\approx 5.8\cdot 10^{10}$. The actual values are 
$\tmax=83$ and $\var{n_{\WT}^\pi(\tmax)}\approx 1.5\cdot 10^{12}$. The somewhat 
odd fact that $\tmax$ is pretty far off, while the variance, which depends very 
strongly on $\tmax$, is reasonably close, is not entirely unexpected: Towards 
the end of the simulation, many $\WT$ molecules will be selected away (as the 
chemical potential for the selection approaches the binding energy of $\WT$), 
which will increase $\tmax$. However, this will also affect the variance 
correspondingly, such that using the solution from equation~\ref{tmax} in 
equation~\ref{MFVarApprox2} will still give the right value. Considering that we 
have used an extremely simple approximation for selection (and completely 
ignored the $\Delta K$ for the various mutations), the estimate seems very 
reasonable. As a check, when using very large energies for the mutations (while 
keeping the ratios fixed), the simulations yield $\tmax=65$ for the mean field 
regime.

\clearpage

\begin{thebibliography}{99}

\bibitem{Liu02}
Bittker, J.A., Phillips, K.J. \& Liu, D.R. (2002) {\it Curr. Opin. Chem.
Biol.} {\bf 6}, 367-374.

\bibitem{RNA}
Landweber, L.F. (1999) {\it Trends Ecol. Evol.} {\bf 14}, 353-358.

\bibitem{Arnold}
Arnold, F.H., ed. (2001) {\it Evolutionary Protein Design} (Academic Press,
San Diego).

\bibitem{Holland}
Novella, I.S., Duarte, E.A., Elena, S.F., Moya, A., Domingo, E. \&
Holland, J.J. (1995) {\it Proc. Natl. Acad. Sci. USA} {\bf 92},
5841-5844.

\bibitem{TsiLevKes}
Tsimring, L., Levine, H. \& Kessler, D.A. (1996) {\it Phys. Rev. Lett.}
{\bf 76}, 4440-4443.

\bibitem{KesLevRidTsi}
Kessler, D.A., Levine, H., Ridgway, D. \& Tsimring, L. (1997) {\it J.
Stats. Phys.} {\bf 87}, 519-544.

\bibitem{RidLevKes}
Ridgway, D., Levine, H. \& Kessler, D.A. (1998) {\it J. Stats. Phys.}
{\bf 90}, 191-210.

\bibitem{PenGerHwaLev}
Peng, W., Gerland, U., Hwa, T. \& Levine, H. (2002) cond-mat/0204117.

\bibitem{Libchaber}                                               
Dubertret, B., Liu, S., Quyang, Q. \& Libchaber, A. (2001) {\it Phys. Rev.
Lett.} {\bf 86}, 6022-6025. 

\bibitem{Stormo97}
Fields, D.S., He, Y., Al-Uzri, A. \& Stormo, G.D. (1997) {\it J. Mol. Biol.}
{\bf 271}, 178-194.

\bibitem{Stormo98}
Stormo, G.D. \& Fields, D.S. (1998) {\it Trends in Biochemical Sciences} 
{\bf 23}, 109-113.

\bibitem{Stormo01}
Man, T.-K. \& Stormo, G.D. (2001) {\it Nucl. Acid. Res.} {\bf 29},
2471-2478. 

\bibitem{NSB1} Record, M.T. Jr., deHaseth, P.L. \& Lohman, T.M. (1977)
{\it Biochemistry} {\bf 16}, 4791-4796.

\bibitem{NSB2} Winter, R.B. \& von Hippel, P.H. (1981) {\it Biochemistry}
{\bf 20}, 6948-6960.

\bibitem{NSB3} Frank, D.E., Saecker, R.M., Bond, J.P., Capp, M.W.,
Tsodikov, O.V., Melcher, S.E., Levandoski, M.M. \& Record, M.T. Jr.
(1997) {\it J. Mol. Biol.} {\bf 267}, 1186-1206.

\bibitem{Peliti}
Peliti, L. (1997) cond-mat/9712027.

\bibitem{WooHig}
Woodcock, G. \& Higgs, P.G. (1996) {\it J. Theor. Biol.} {\bf 179},
61-73.

%\bibitem{HipBer}
%P.H. von Hippel and O.G. Berg, {\em Proc. Natl. Acad. Sci.} {\bf 83},
%1608 (1986); O.G. Berg and P.H. von Hippel, {\em J. Mol. Biol.} {\bf 193},
%723 (1987).

%\bibitem{BruDer}
%E. Brunet and B. Derrida, {\em Phys. Rev. E} {\bf 56}, 2597 (1997).

%\bibitem{PecLev}
%L. Pechenik and H. Levine, {\em Phys. Rev. E} {\bf 59}, 3893 (1999).

\bibitem{Kloster}
Kloster, M. unpublished.

\bibitem{MutRateRef1}
Shafikhani, S., Siegel, R., Ferrari, E. \& Schellenberger, V. (1997)
{\it Biotechniques} {\bf 23}, 304-310.

\bibitem{MutRateRef2}
Cadwell, R. \& Joyce, G. (1992) {\it PCR Meth. Appl.} {\bf 2}, 28-33.

\bibitem{MutRateRef3}
Lin-Goerke, J., Robbins, D. \& Burczak, J. (1997)
{\it Biotechniques} {\bf 23}, 409-412.

\end{thebibliography}
\end{document}